# A Two-Step Framework for Multi-Material Decomposition of Dual Energy Computed Tomography from Projection Domain

Di Xu[1], Qihui Lyu[1], Dan Ruan[2] and Ke Sheng[1,*]

1 Department of Radiation Oncology, University of California, San Francisco

2 Department of Radiation Oncology, University of California, Los Angeles

{di.xu; ke.sheng}@ucsf.edu

## Abstract

*Background and Purpose:* Dual-energy computed tomography (DECT) utilizes separate X-ray energy spectra to improve multi-material decomposition (MMD) for various diagnostic applications. However accurate decomposing more than two types of material remains challenging using conventional methods. Deep learning (DL) methods have shown promise to improve the MMD performance, but typical approaches of conducing DL-MMD in the image domain fail to fully utilize projection information or under iterative setup are computationally inefficient in both training and prediction. In this work, we present a clinical-applicable MMD (>2) framework rFast-MMDNet, operating with raw projection data in non-recursive setup, for breast tissue differentiation.

*Methods*: rFast-MMDNet is a two-stage algorithm, including stage-one SinoNet to perform dual energy projection decomposition on tissue sinograms and stage-two FBP-DenoiseNet to perform domain adaptation and image post-processing. rFast-MMDNet was tested on a 2022 DL-Spectral-Challenge dataset, which includes 1000 pairs of training, 10 pairs of validation, and 100 pairs of testing images simulating dual energy fast kVp-switching fan beam CT projections of breast phantoms. MMD for breast fibroglandular, adipose tissues and calcification was performed. The two stages of rFast-MMDNet were evaluated separately and then compared with four noniterative reference methods including a direct inversion method (AA-MMD), an image domain DL method (ID-UNet), AA-MMD/ID-UNet + DenoiseNet and a sinogram domain DL method (Triple-CBCT).

*Results*: Our results show that models trained from information stored in DE transmission domain can yield high-fidelity decomposition of the adipose, calcification, and fibroglandular materials with averaged RMSE, MAE, negative PSNR, and SSIM of 0.004±~0, 0.001±~0, -45.027±~0.542, and 0.002±~0 benchmarking to the ground truth, respectively. Training of entire rFast-MMDNet on a 4×RTX A6000 GPU cluster took a day with inference time <1s. All DL methods generally led to more accurate MMD than AA-MMD. rFast-MMDNet outperformed Triple-CBCT, but both are superior to the image-domain based methods.

*Conclusions*: A fast, robust, intuitive, and interpretable workflow is presented to facilitate an efficient and precise MMD with input from projection domain.

## 1. Introduction

Computed tomography (CT) imaging has been an indispensable tool for non-invasive diagnosis since its invention in the 1970s[1]. The contrast of CT is determined by a combination of X-ray material interactions, including photoelectric and Compton, while the CT number primarily reflects electron density with a component dependent on the material composition. Conventional CT is obtained using a single source detector without explicit energy separation in the single energy CT (SECT) mode[2]. However, the information preserved in SECT is inadequate to resolve the individual subcomponents contributing to the CT number. In pursuing better material differentiation and contrast, dual-energy CT (DECT) was invented[3] to acquire attenuation information with different energies or energy spectra. Several different mechanisms have been used to acquire DECT, including X-Ray tubes operating at different voltages[4], filters to modify the X-RAY spectra[2], detectors with different sensitivities to X-RAY energies[5,6], and photon-counting detectors[7]. DECT has shown promise in various clinical applications, including iodine mapping[8], gout classification[9], fat and essential trace metal quantification[10,11], bone separation [12], and virtual unenhanced and monochromatic image generation[13].

While DECT offers additional information for multi-material decomposition (MMD), the problem itself is still non-linear and ill-posed, particularly for separating $> 2$ materials[14,15]. Additional priors of the materials are often assumed to achieve acceptable decomposition results. Existing MMD ($\geq 2$) typically follows two approaches: theory-integrated methods that conduct decomposition with the assistance of a physical model[16], or analytical methods that perform direct attenuation decomposition in dual-energy (DE) image or projection domain[16,17(p)]. Among them, the theory-integrated algorithms and analytical methods directly process projections and are thus robust to beam hardening artifacts[16,17]. Theory-integrated techniques requiring iterative operations of forward and backward projections are extremely computationally expensive [16]. Projection-domain MMD has been conditioned for 2 basis materials[17] that are inadequate for more complex patient MMD problems. Though analytical methods based on image domain have been demonstrated for >2 material decomposition, the decomposed results highly depend on strong priors, which limit its generalizability, and are susceptible to image quality[18–21].

In the past decade, as new annotated data and powerful computational resources emerged, deep neural networks (DNN) started to realize their potential for learning complex relationships and incorporating existing knowledge into the inference model[22]. Multiple studies have demonstrated the advantage of DNN-inspired DECT reconstruction models for fast MMD while suppressing noise and artifacts[23–25]. For instance, Zhang et al. proposed a butterfly network to acquire image-domain decomposition of two basis materials[26]. A visual geometry group (VGG) network with an enlarged receptive field was trained by Chen et al. to perform MMD on five simulated basis materials[27]. Badea et al. deployed U-Net for the tasks of prediction in the photoelectric effect, Compton scattering, and material density maps[28]. For more related works, we refer the audience to[29–31]. Nevertheless, these algorithms performed MMD in the image domain suffered from information loss caused by converting DE transmission to image and were suboptimal with the data-driven nature of DL methods. They were limited in learning efficiency and MMD performance for challenging problems, including the separation of sparse structures such as calcification from noise and material boundary definition between tissues with similar atomic compositions.

Lately, Zhu et al. performed DNN-based MMD of more than two basis materials directly from the dual energy projections[32]. Specifically, they introduced a framework termed Triple-CBCT, which consisted of a cascade of three sub-networks: 1) SD-Net - a seven-layer convolutional neural network (CNN) with sparse skip-connection and mixed sizes of the convolutional filter designed to generate the sinograms of numerous virtual monochromatic images (VMI) from the domain of DE transmission; 2) Domain transform Net - a fix-parameter predictor with the integration of standard Feldkamp-Davis-Kress image reconstruction algorithm[33] to map sinogram signals to image domain; 3) ID-Net - an eight-layer CNN with dense inter-layer skip-connections and 3x3 convolutional filter for denoising image, suppressing artifact- and grouping multiple VMIs generated from 2) to the final base material images. Triple-CBCT shows feasibility of performing >2 MMD from DE projections. Yet, the downsides of Triple-CBCT are its excessive and uninterpretable network architectures and heavy training parameters.

More recently, Sidky et al. reported the top-ranked algorithms in the 2022 AAPM DL-spectral-challenge[34], where most pipelines decomposed materials from the sinogram domain and achieved remarkably low root mean square error (RMSE). However, they are limited to offline applications due to the dependence on conventional and slow iterative reconstruction as a sub- or main component. The iterative forward and backward projections are extremely computationally expensive in training and testing stages and incapable of generating real-time predictions. For instance, Hauptmann et al. reported that an iterative DL model (5 forward and backward iterations) are around five times slower than non-iterative learning methods[35]. Similar comparison were reported by other authors [36,37].

To this end, we believe that the encouraging statistical results from Zhu et al.[32] and the 2022 AAPM challenge[34] warrant further development for improvement in model applicability. Specifically,  a workflow using a lightweight and interpretable network structure is desirable without losing quality control (QC). We realize such a vision in the novel rFast-MMDNet, which is elaborated as follows.

The proposed rFast-MMDNet includes 1) SinoNet – a UNet[38] based architecture for the decomposition of DE projection to basis material sinograms; 2) DenoiseNet – a customized denoising CNN consisting of sixteen ResNet[39] blocks and an integrated one-time filter back projection (FBP)[40] for fast conversion of signals from sinogram to image. Notably, the training of SinoNet and DenoiseNet is conducted separately for fast and stable convergence of feature extraction in each stage.  The rest of our manuscript is organized as follows: **Sec. 2** presents the design and training of rFast-MMDNet, **Sec. 3** elaborates on visual and statistical results, and **Sec. 4 and 5** discuss and summarizes this work.

## 2. Method

### 2.1 Dataset

#### 2.1.1   Simulated Imaging Cohort

The 2022 American Association of Physicists in Medicine (AAPM) DL-spectral-CT challenge released a set of simulated DECT scans with ground truth (GT) tissue phantoms for a breast model[41] containing three materials: adipose, fibroglandular, and calcification (assumed to be composed of hydroxyapatite[42]). The DL-spectral-CT dataset modeled ideal fast kVp-switching (50 and 80 kVps) acquisition of circular fan-beam X-rays projected on a flat panel detector with 1024 pixels. The 50-/80-kVp transmission ($I_w$) is generated under the assumption of a spectral CT model formulated as **Equation (1)** with assumed spectral sensitivities ($S$) and linear attenuation ($\mu_m$) distributed as the upper row of **Fig. 1**. The dataset has 1000 pairs of training, 10 pairs of validation, and 100 pairs of testing images. Each of the inputs of low(50)/high(80)-kVp acquisitions has 256 projection views. The GTs are all 2-dimensional (2D) and have a height (H) × width (W) of $512 \times 512$. Additionally, both the input and GT signals are normalized to between 0-1.  The bottom row of **Fig. 1** shows the data processing pipeline.

$$I_w = \int S_w(E) \exp\left[-\mu_a(E)Px_a - \mu_f(E)Px_f - \mu_c(E)Px_c\right] dE \quad (1)$$

Where $w$ index is either 50 or 80 kVp, $m$ index is ether adipose ($a$), fibroglandular ($f$), or calcification ($c$), $x_m$ is tissue phantom, and $P$ is short for forward projection.

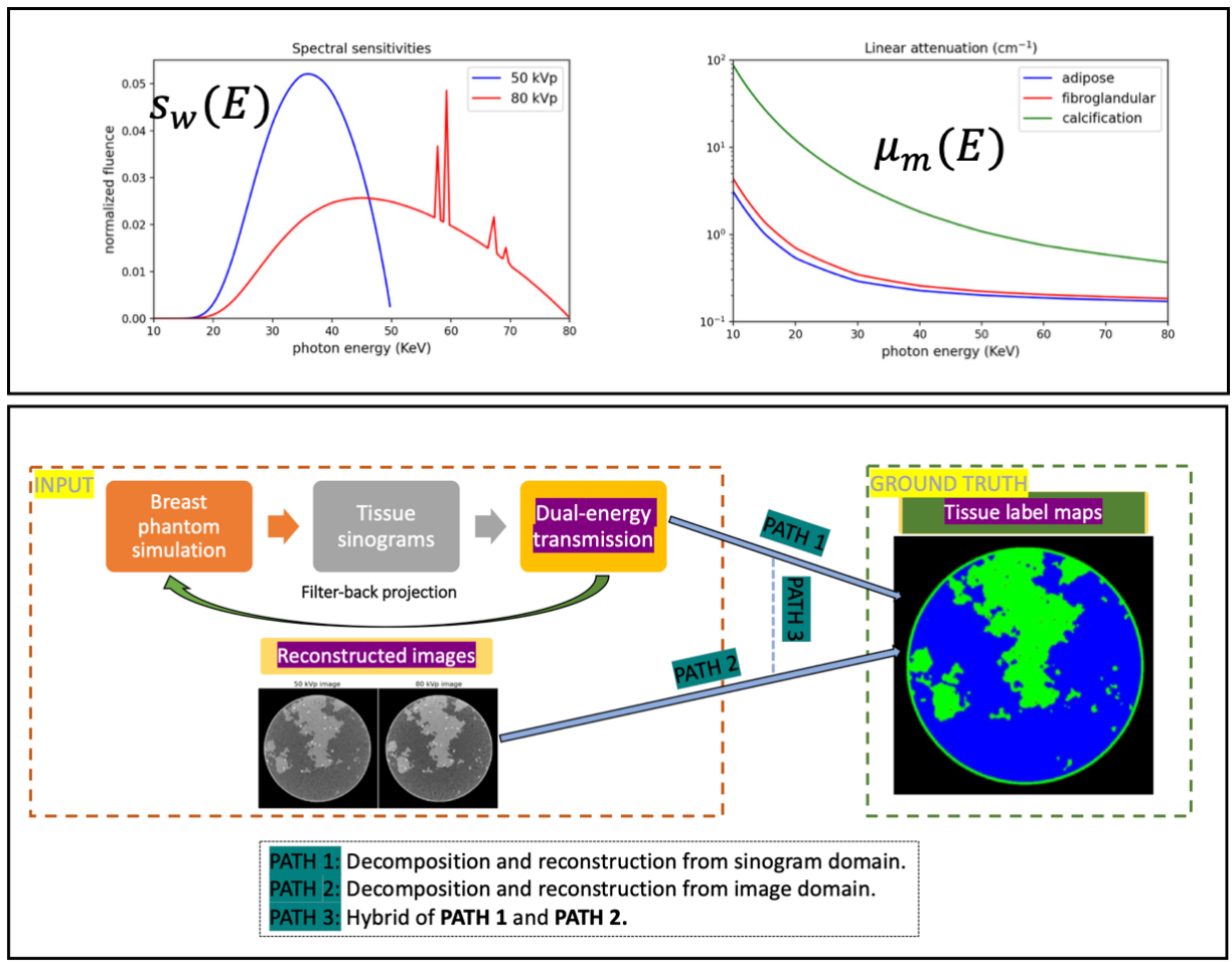

**Fig. 1**: The upper row shows the spectral sensitivity and linear attenuation distribution utilized for simulating the low-/high-energy transmission. The bottom row is a flowchart describing the simulation process of the DL-Spectral-CT dataset. The 50/80 kVp images on the bottom left are reconstructed from DE transmissions using FBP for visualization purposes. The image on the right-hand side shows the label map of GT with adipose in blue, fibroglandular in green, and microcalcification in red.

### 2.1.2  *Data Preprocessing for Model Training*

Regarding the input of SinoNet, we first created two zero-filling $1 \times 512 \times 1024$ matrices named zero-matrix-high and zero-matrix-low, respectively. To process the high-kVp transmission, we arranged the sequence from each view of a raw projection into the odd rows of zero-matrix-high with no modification in the even rows. Vice versa for low-KVp, for which we placed the projections in the even rows of zero-matrix-low and left odd rows unchanged. The matrix structure reproduced the alternating pattern of fast kVp switching low- and high-energy X-ray data collection. Next, we stacked the dilated 2D high- and low-kVp signals channel-wise to form the input volume with the dimension of Channel (C) $\times$ H $\times$ W in $2 \times 512 \times 1024$. To process GT, the forward projection was performed first to generate separate sinograms for individual materials in the phantom, and then the forward projection sinograms were channel-wise concatenated for the three basis materials with a dimension of C $\times$ H $\times$ W = $3 \times 512 \times 1024$. Lastly, resizing (scales in $[0.7, 0.9, 1.0, 1.1, 1.2]$) was applied to augment the training data five-fold while perserving the projection geometry within the input volume.

DenoiseNet combined both input (FBP images) and GT tissue phantoms to form a new matrix ( C $\times$ H $\times$ W = $3 \times 512 \times 512$ ). Data augmentation, including random rotation (angles in $[-30°, -15°, 0°, 15°, 30°]$), random resizing (scales in $[0.7, 0.9, 1.0, 1.1, 1.2]$ ), random gaussian blur (kernels in $[1, 3, 5]$), random cropping, random brightness, and mirroring was implemented.

## 2.2 rFast-MMDNet Model

**Fig. 2** shows the complete architecture of rFast-MMDNet. As mentioned in **Sec. 1**, the overall pipeline is divided into two stages. In the first stage, SinoNet learns decomposed raw DE projection features in the sinograms. In the second stage, rough tissue patterns are generated via FBP-based domain adaptation, further refined by a DenoiseNet to suppress noise and remove artifacts. Detailed illustrations and justification regarding each module are elucidated as follows.

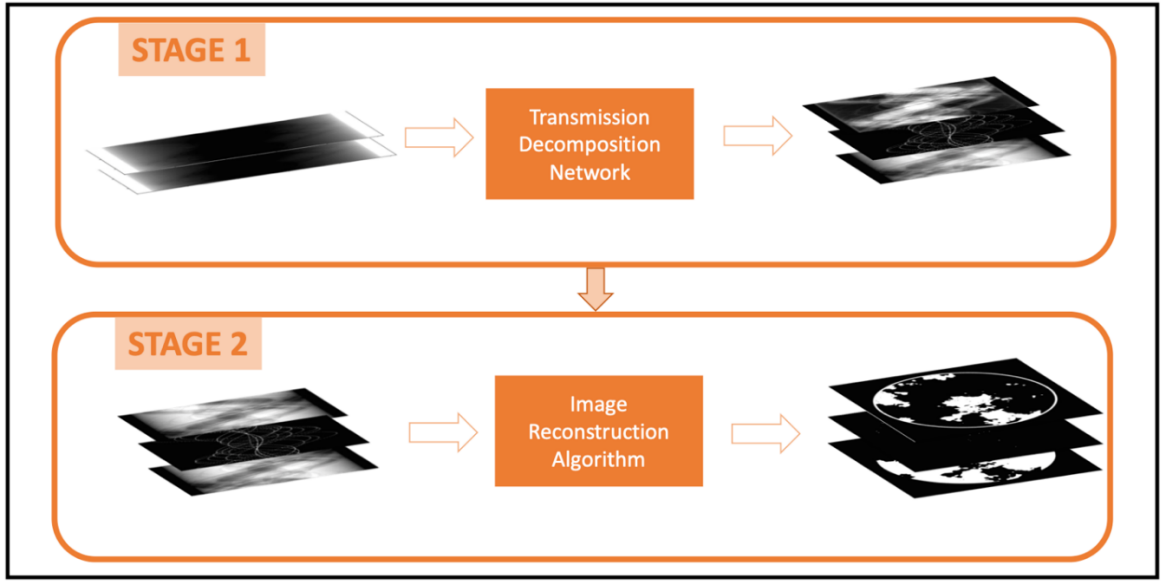

**Fig. 2**: The complete pipeline of rFast-MMDNet. The upper row (stage 1) works on transmission decomposition and the bottom row (stage 2) works on FBP-based domain adaptation – sinogram -> image, then followed by DenoiseNet for denoising and artifact-removal of FBP images.

### 2.2.1   SinoNet

**Fig.3** shows the structure of SinoNet for DE projection decomposition. In particular, SinoNet was inspired by Xu et al.[43] designed using a 3D UNet-based encoder-decoder combined with skipped information feeding from each encoding to its corresponding decoding layer via channel-wise concatenation. The $3 \times 3$ convolution and $2 \times 2$ max pooling filters were applied to all layers in SinoNet. Altogether, SinoNet has one bottleneck block for structure balancing and 7 separate encoding and decoding layers. The number of spatial filter channels doubles per layer from the base of 32 to a maximum of 320, and the feature map dimension halves per individual step of encoding convolution. Vice versa, the decoding blocks reduce the filter channel number and increase the figure map dimension by 2. The rationale of such a network structure can be understood as follows. The encoding stage casts the DE transmissions to a higher dimension depicting various VMI projections at distinct energy levels. The decoding stage then combines virtual monochromatic projections into the sinograms of adipose, fibroglandular, and calcification, respectively. The architecture reduces the network complexity, facilitates efficient training convergence, and maintains model robustness. As discussed in **Sec. 2.1.2**, the input of SinoNet was dilated and channel-wise concatenated DE projections, while the corresponding GTs were channel-wise stacked sinograms forward-projected from tissue phantoms.

The residual discrepancy between the SinoNet predicted sinograms and the theoretical material sinograms was minimized using DenoiseNet in the second stage.

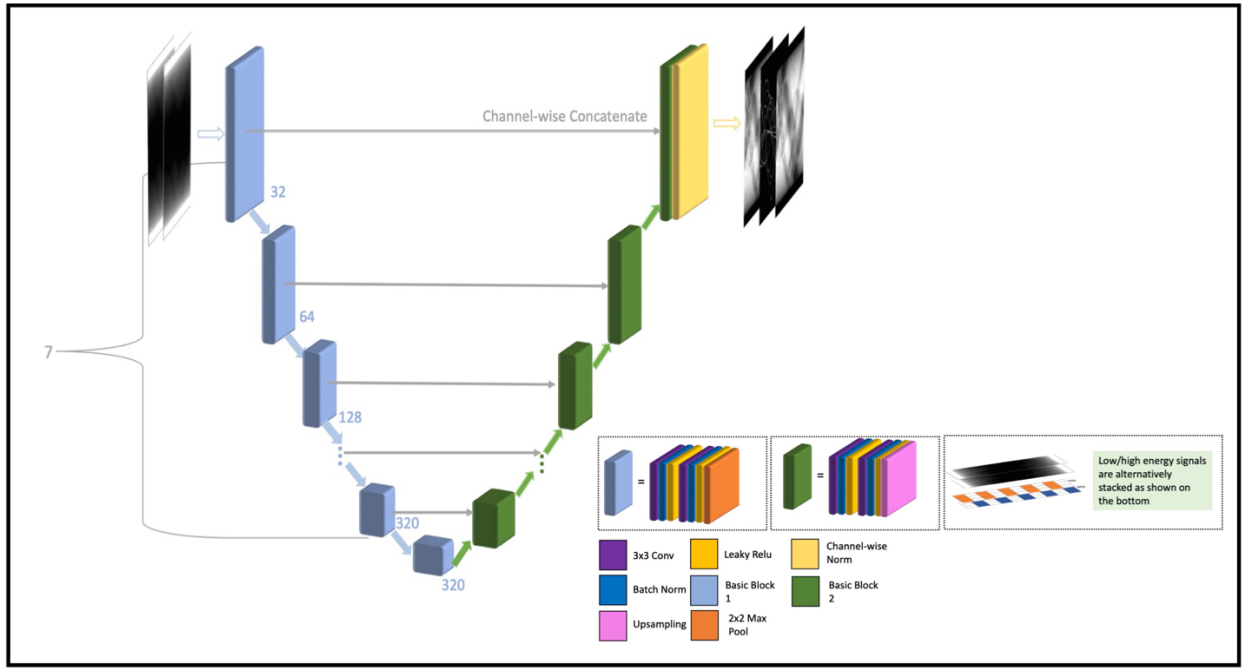

**Fig. 3**: architecture of SinoNet for decomposing material sinograms from DE transmission signals. The blocks in blue represent the encoding layer while those in green represent the decoding layer.

### 2.2.2   FBP + DenoiseNet

As demonstrated in the first row of **Fig. 4**, the image reconstruction scheme entails two sub-components: 1) FBP image generator to characterize the coarse morphology of tissue phantoms in the image domain; 2) DenoiseNet for detail refinement. DenoiseNet (second row in **Fig. 4**) consists of 16 standard ResNet blocks (structure shown in the third row of **Fig. 4**), each having a convolution kernel of size $3 \times 3$, a stride of $1$, and spatial filter channels in an increase-to-decrease form. Pooling layers are not applied in DenoiseNet to maintain the resolution of feature maps.

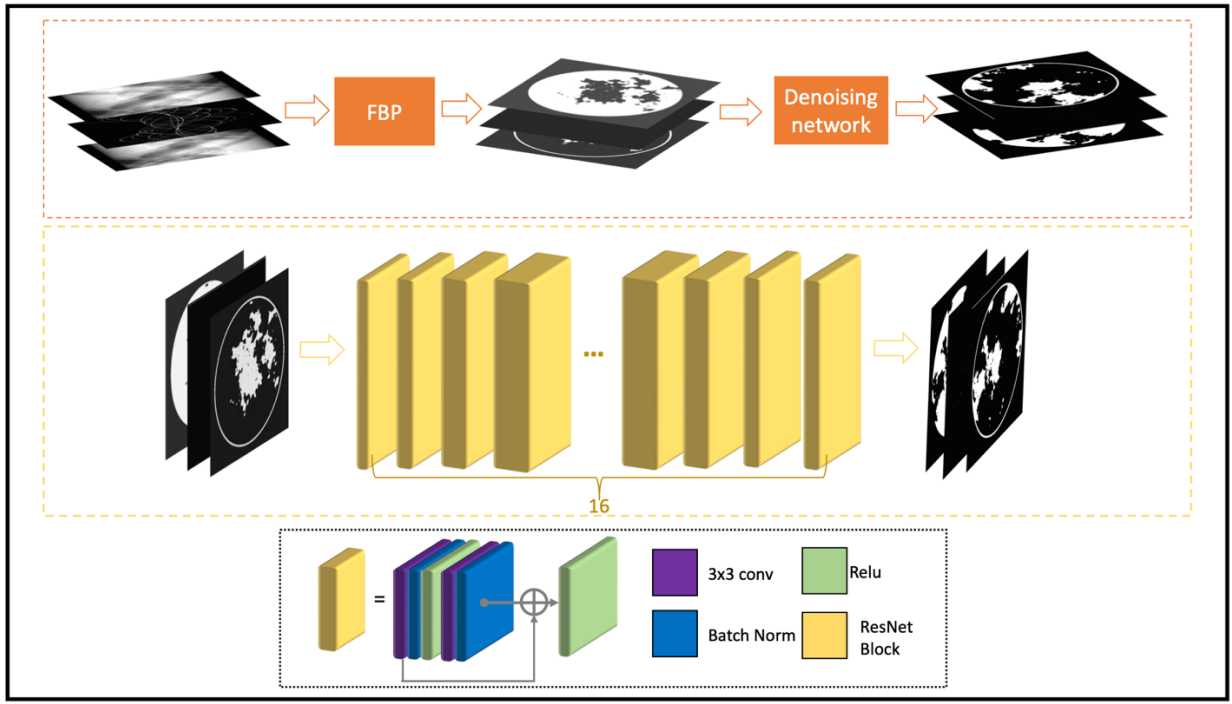

**Fig. 4**: The first row presents the generic workflow of stage two for accurate image reconstruction. The second row visualizes the architecture of DenoiseNet. And the last row illustrates the composition of each ResNet block of our DenoiseNet.

### 2.2.3  Loss Function

SinoNet and DenoiseNet share the same cost function combining negative peak signal-to-noise ratio (PSNR), multi-scale structure similarity index measure (MS-SSIM), and $L_1$ deviation measurement:

$$L = -\alpha \cdot L_{PSNR} + (1-\alpha) \cdot [\beta \cdot (1 - L_{MS-SSIM}) + (1-\beta) \cdot L_1] \quad (2)$$

$$L_{PSNR} = log_{10}\left(\frac{MAX_x^2}{1/mn \cdot \sum_{i=0}^{m-1}\sum_{j=0}^{n-1}[x_{ij} - y_{ij}]^2}\right) \quad (3)$$

$$L_{MS-SSIM} = \frac{(2\mu_x\mu_y + c_1)}{(\mu_x^2 + \mu_y^2 + c_1)} \cdot \prod_{j=1}^{M}\frac{(2\sigma_{x_jy_j} + c_2)}{(\sigma_{x_j}^2 + \sigma_{y_j}^2 + c_2)} \quad (4)$$

$$L_1 = \frac{1}{mn} \cdot \sum_{i=0}^{m-1} \sum_{j=0}^{n-1} ||x_{ij} - y_{ij}||_1 \tag{5}$$

Where $\alpha$ and $\beta$ in **Equation (2)** are hyperparameters set to 0.5 and 0.25, respectively, $X$ and $Y$ are model input and target, $MAX_X$ is the maximum possible input value, $[\sigma_{x_j y_j}, \dots, \sigma_{x_M y_M}]$ of **Equation (3)** are set to $[0.5, 1.0, 2.0, 4.0, 8.0]$ , $c_1 = (k_1 S)^2$ and $c_2 = (k_2 S)^2$ of **Equation (4)** are two variables to stabilize the division with a weak denominator $S$ as the dynamic range of the pixel-values (typically $2^{\# \, bits \, per \, pixel} - 1$), $(k_1, k_2)$ are constants, and $||\cdot||_1$ denotes the $l_1$ norm.

### 2.2.4 Model Training

Both SinoNet and DenoiseNet were implemented in PyTorch. The training was performed on a GPU cluster with $4 \times$ RTX A6000.

For SinoNet training, we set the maximum number of epochs as 1k and observed the model convergence at $\sim$450 epochs. Adam optimizer with an initial learning rate (LR) of 0.001 and batch size of $4 \times 2$ was applied during learning.

Regarding training of DenoiseNet, under the assumption that limited noises and artifacts exist in the FBPed images, we set the max-epoch number as 200 and found that the validation loss reached plateau at $\sim$70 epochs. Adam optimizer with an initial learning rate (LR) of 0.01 and batch size of $4 \times 4$ was implemented during model tuning.

It is worth noting that all the data augmentations described in **Sec. 2.1.2** were performed on the fly along with DNN training.

## 2.3 Benchmark Algorithms

We compare our in-house rFast-MMDNet with representative MMD methods in four distinct categories. 1) Classical image-domain MMD: We selected an analytical algorithm MMD (AA-MMD) based on the assumption of volume conservation $\leq 3$ materials in an individual voxel. For each voxel, AA-MMD loops over a material triplet library, identify the best-fit triplet, and perform MMD via direct matrix inversion[44]; 2) DL image-domain MMD: We chose a UNet-based image-domain MMD (ID-UNet), which feeds the low- and high-energy images into a 3D UNet and predicts out decomposed materials[28]; 4) image-domain MMD followed by in-house DenoiseNet: to fairly compare the overall efficacy of image and our in-house two-stage pipeline, we also build up two algorithms which run our proposed DenoiseNet after AA-MMD and ID-UNet; 4) DL projection-domain MMD: We selected the previously mentioned Triple-CBCT[32].

## 3. Results

The results from SinoNet and our rFast-MMDNet are reported. For performance comparison, the outcomes from the benchmark algorithms (AA-MMD, ID-UNet, Triple-CBCT) along with the intermediate results from our in-house algorithm (SinoNet + FBP) and that from the rFast-MMDNet pipeline are evaluated. The effectiveness of our DenoiseNet post-processing can be seen from the SinoNet + FBP results. Deviation metrics, including RMSE and mean absolute error (MAE), and similarity measurements, including negative PSNR and SSIM, are incorporated as evaluation metrics. Both quantitative and qualitative outcomes are reported.

## 3.1 Results of SinoNet

As presented in **Fig. 5**, GT and SinoNet results are visually indistinguishable. Quantitative results also show nearly negligible deviation from GT with validation RMSE, MAE, negative PSNR, and SSIM of $0.002 \pm \sim0$, $0.001 \pm \sim0$, $-55.738 \pm 0.314$, and $0.001 \pm \sim0$, and test of $0.002 \pm \sim0$, $0.001 \pm \sim0$, $-54.872 \pm 0.347$, and $0.001 \pm \sim0$ of averaged mean values among three tissues.

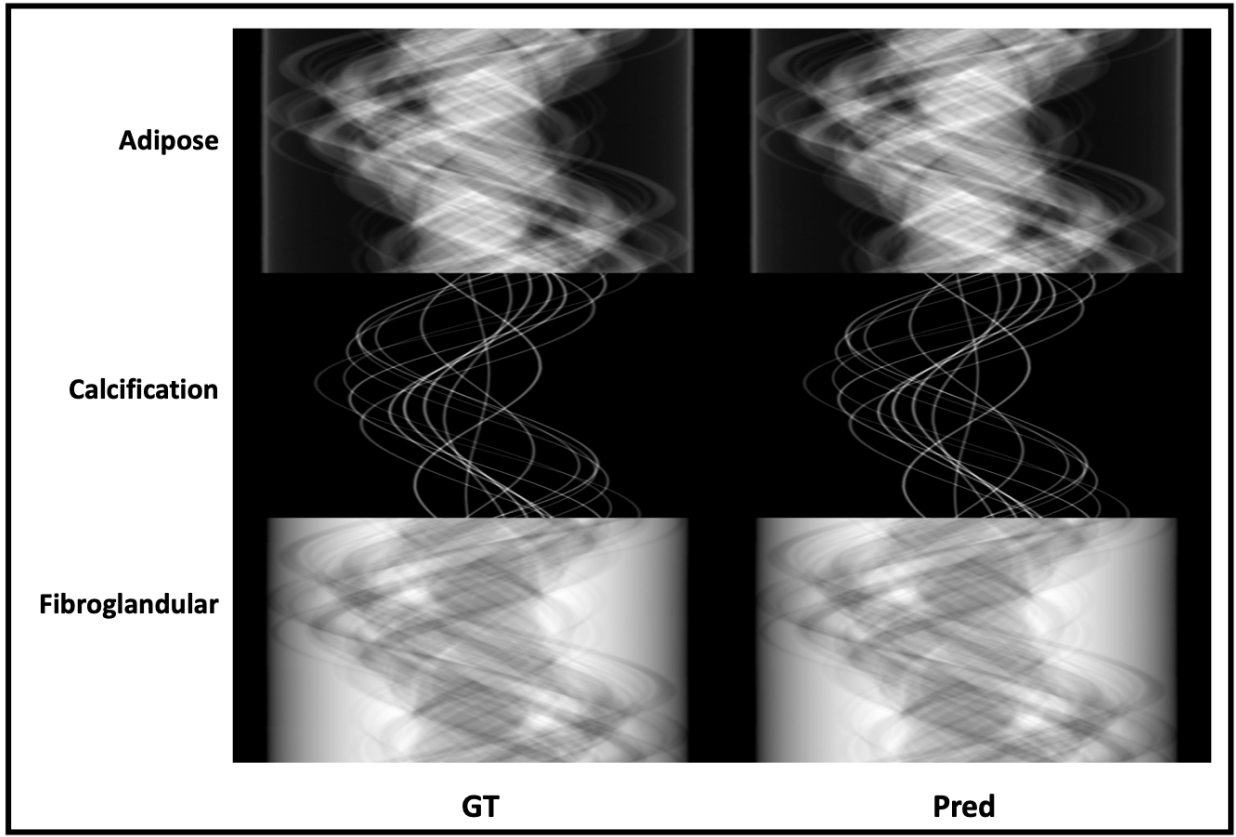

**Fig. 5**: Left-hand side shows the GTs from forward projection phantom sinograms and right-hand side shows the predictions of fully trained SinoNet in a test case. Tissues of adipose, calcification and fibroglandular are listed from top to bottom row-wise.

### 3.2 Results of rFast-MMDNet

In **Fig. 6**, we can observe that the predictions from AA-MMD show visible losses of information, including poor image resolution, incorrect boundaries of subtle structures (e.g., adipose and fibroglandular), and noisy predictions of calcification. ID-UNet mitigates the noise in calcification predictions and is slightly better at preserving anatomical details but still shows discrepancies of inaccurate micro-pattern delineation with GT (see the visualization of ID-UNet in **Fig. 6**). Next, SinoNet+FBP can roughly characterize the anatomical components but is quantitatively inaccurate due to beam hardening artifacts. Finally, we found that Triple-CBCT and our rFast-MMDNet are qualitatively best performers, which produce visually similar decomposition results.

Quantitatively, we can see from **Tab. 1** and **Tab. 2** that rFast-MMDNet outperforms Triple-CBCT in all cases, showing $(0.001, 0.003)$ lower error metrics and $\sim 0.006$ and $\sim -3$ better in similarity metrics. Both rFast-MMDNet and Triple-CBCT are markedly better than AA-MMD and SinoNet-FBP by two orders of magnitude in deviation from GT and $(10, 30)$ in negative PSNR. Additionally, even though running DenoiseNet on top of AA-MMD and ID-UNet, image-domain MMD algorithms are still far worse than sinogram domain MMD. Besides mean values, rFast-MMDNet has the lowest standard deviation (std) $\sim 0$ standard deviation in RMSE, MAE, and SSIM, and 0.453/0.542 in negative PSNR for both validation and test sets in **Tab. 1**. The observation is consistent with individual materials analysis In **Tab. 2**. It is worth noting that the calcification prediction errors for all MMD methods are lower than adipose and fibroglandular tissue predictions, likely due to its high contrast and sparse representation in the phantoms.

**Tab. 1** shows that all image domain methods make sub-second predictions. In comparison, Triple-CBCT is significantly slower, taking up to 174 seconds to finish a decomposition prediction. Our proposed rFast-MMDNet achieved comparable or faster speed with image-domain methods while maintaining the advantages of being a sinogram-domain method. Among all DL models, Triple-CBCT has the heaviest model parameters and computation demand and image domain algorithms have roughly half of the model parameters and MACs than sinogram domain methods.

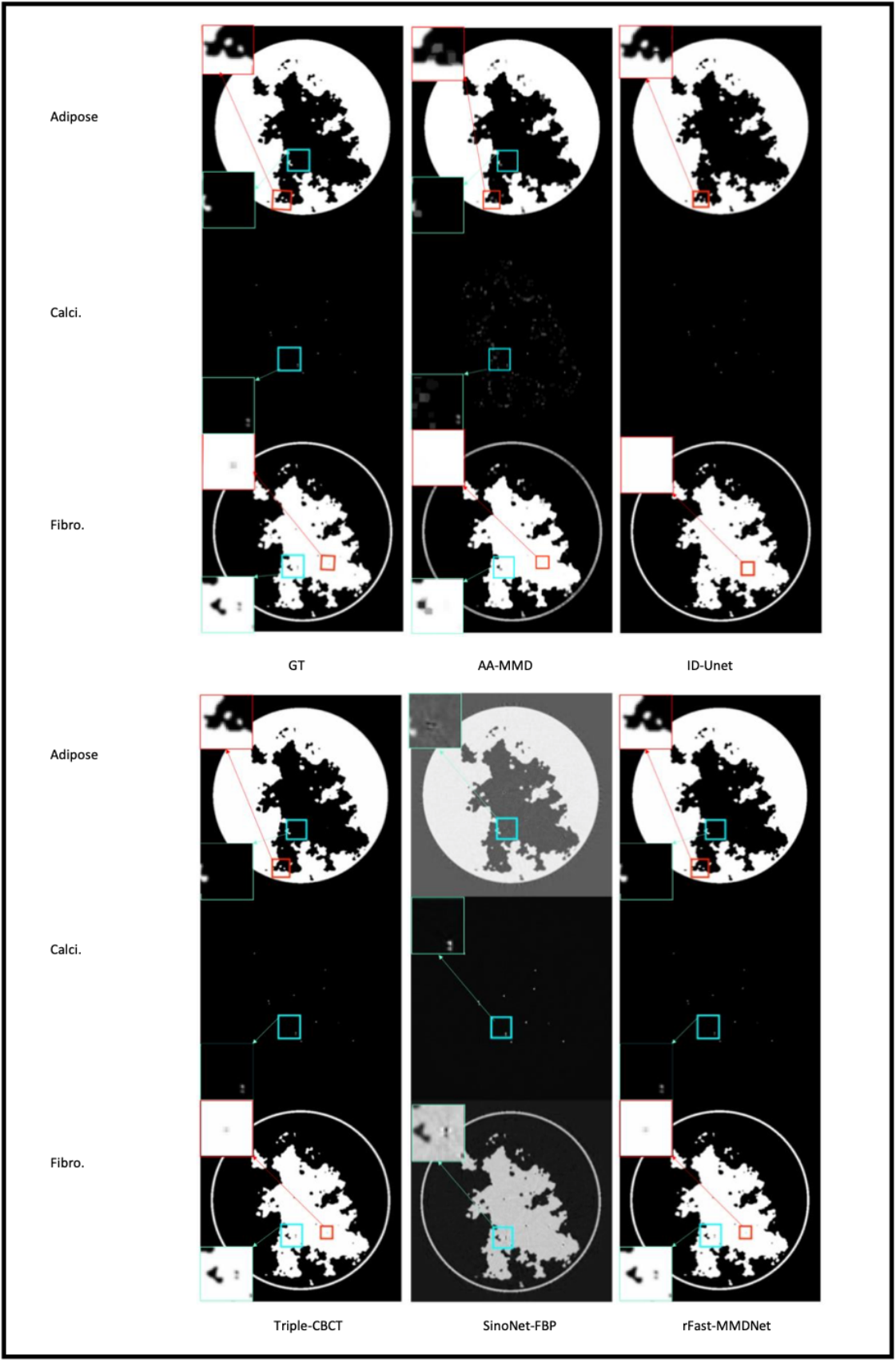

**Fig. 6**: GT and MMD predictions from benchmark and in-house algorithms of Adipose, calcification (calci.), and fibroglandular (fibro.) on a test case. Red/blue bounding boxes on the upper-right/lower-right shows zoomed-in predictions from different algorithms on the same region.

| Domain | Dataset | Method | RMSE | MAE | -PSNR | SSIM | EXEC (ms) | Parameters (M) | MACs (G) |
|---|---|---|---|---|---|---|---|---|---|
| Image | Validation | AA-MMD[44] | 0.195 ± 0.107 | 0.186 ± 0.014 | −13.453 ± 1.874 | 0.472 ± 0.003 | | | |
| | | AA-MMD[44]+DenoiseNet | 0.023±0.007 | 0.031±0.012 | −24.335±0.945 | 0.023±0.042 | | - | |
| | | ID-UNet[28] | 0.013 ± 0.003 | 0.015 ± 0.017 | −27.994 ± 0.731 | 0.021 ± 0.027 | | | |
| | | ID-UNet[28] + DenoiseNet | 0.010±0.003 | 0.011±0.009 | −36.732±0.631 | 0.009±0.004 | | | |
| | Test | AA-MMD[44] | 0.183 ± 0.112 | 0.175 ± 0.019 | −15.774 ± 1.457 | 0.421 ± 0.004 | 204 | - | - |
| | | AA-MMD[44] + DenoiseNet | 0.021±0.006 | 0.028±0.011 | −26.543±0.902 | 0.022±0.039 | 291 | 44.790 | 67.520 |
| | | ID-UNet[28] | 0.010 ± 0.004 | 0.012 ± 0.020 | −29.827 ± 0.871 | 0.018 ± 0.020 | **67** | 35.210 | 60.330 |
| | | ID-UNet[28] + DenoiseNet | 0.009±0.003 | 0.010±0.010 | −37.374±0.624 | 0.008±0.003 | 150 | 80.000 | 80.000 |
| Projection | Validation | Triple-CBCT[33] | 0.011 ± 0.005 | 0.0106 ± 0.012 | −35.143 ± 0.532 | 0.010 ± ~0 | | | |
| | | SinoNet-FBP | 0.195 ± 0.072 | 0.134 ± 0.011 | −13.717 ± 1.286 | 0.410 ± 0.003 | | - | |
| | | rFast-MMDNet | **0.005 ± ~0** | **0.002 ± ~0** | **−43.792 ± 0.453** | **0.004 ± ~0** | | | |
| | Test | Triple-CBCT[32] | 0.007 ± 0.006 | 0.008 ± 0.015 | −38.571 ± 0.631 | 0.008 ± ~0 | 174000 | 118.590 | 147.030 |
| | | SinoNet-FBP | 0.173 ± 0.095 | 0.120 ± 0.095 | −15.616 ± 1.378 | 0.381 ± 0.003 | 101 | **32.870** | **57.360** |
| | | rFast-MMDNet | **0.004 ± ~0** | **0.005 ± ~0** | **−45.027 ± 0.542** | **0.002 ± ~0** | 184 | 77.660 | 124.880 |

**Tab. 1**: The measured mean values and stds of difference between GTs and predictions among all tissue phantoms. Prediction execution time, model parameters and Multiply–accumulate operations (MACs) are also reported. model Values marked with wavy underline are the weakest scores in the validation set, those marked in straight underline are the weakest scores in test set. The strongest scores in validation and test sets are denoted in bold fonts.

| Material | Dataset | Method | RMSE | MAE | -PSNR | SSIM |
|---|---|---|---|---|---|---|
| *Adipose* | Validation | AA-MMD[44] | <u>0.207 ± ~0</u> | <u>0.189 ± ~0</u> | <u>−12.712 ± 0.779</u> | <u>0.489 ± 0.004</u> |
| | | AA-MMD[44] + DenoiseNet | 0.032 ±~0 | 0.033 ±~0 | −21.702±0.873 | 0.005±0.023 |
| | | ID-UNet[28] | 0.015 ± ~0 | 0.019 ± ~0 | −30.829 ± 0.612 | 0.003 ± 0.018 |
| | | ID-UNet[28] +DenoiseNet | 0.012±~0 | 0.017 ± ~0 | −32.801 ± 0.617 | 0.005 ± 0.019 |
| | | Triple-CBCT[32] | 0.012 ± ~0 | 0.004 ± ~0 | −33.345 ± 0.491 | 0.016 ± 0.006 |
| | | SinoNet-FBP | 0.137 ± 0.003 | 0.128 ± 0.005 | −16.153 ± 0.532 | 0.017 ± 0.007 |
| | | rFast-MMDNet | **0.008 ±~0** | **0.003 ±~0** | **−40.134 ± 0.490** | **0.011 ± 0.002** |
| | Test | AA-MMD[44] | <u>0.197 ± ~0</u> | <u>0.178 ± ~0</u> | <u>−13.224 ± 0.892</u> | <u>0.471 ± 0.005</u> |
| | | AA-MMD[44] + DenoiseNet | 0.029 ±~0 | 0.032 ±~0 | −22.515±0.794 | 0.004±0.019 |
| | | ID-UNet[28] | 0.013 ± ~0 | 0.015 ± ~0 | −32.187 ± 0.781 | 0.002 ± 0.021 |
| | | ID-UNet[28] + DenoiseNet | 0.011 ± ~0 | 0.013 ± ~0 | −33.072 ± 0.604 | 0.005 ± 0.016 |
| | | Triple-CBCT[32] | 0.010 ± ~0 | 0.004 ± ~0 | −35.271 ± 0.562 | 0.013 ± 0.009 |
| | | SinoNet-FBP | 0.136 ± 0.003 | 0.123 ± 0.008 | −17.402 ± 0.605 | 0.013 ± 0.009 |
| | | rFast-MMDNet | **0.007 ±~0** | **0.002 ±~0** | **−43.193 ± 0.657** | **0.008 ± 0.003** |
| *Calcification* | Validation | AA-MMD[44] | <u>0.084 ± ~0</u> | <u>0.061 ± ~0</u> | <u>−18.893 ± 0.412</u> | <u>0.572 ± 0.004</u> |
| | | AA-MMD[44] + DenoiseNet | 0.035±~0 | 0.010±~0 | −27.468±0.573 | 0.012±0.001 |
| | | ID-UNet[28] | 0.011 ± ~0 | 0.007 ± ~0 | −30.548 ± 0.459 | 0.004 ± ~0 |
| | | ID-UNet[28] +DenoiseNet | 0.011 ± ~0 | 0.006 ± ~0 | −31.764 ± 0.394 | 0.003 ± ~0 |
| | | Triple-CBCT[32] | 0.007 ± ~0 | 0.003 ± ~0 | −38.893 ± 0.391 | 0.005 ± ~0 |
| | | SinoNet-FBP | 0.043 ± ~0 | 0.041 ± 0.001 | −26.879 ± 0.535 | 0.510 ± 0.001 |
| | | rFast-MMDNet | **0.003 ±~0** | **0.001 ± ~0** | **−53.658 ± 0.732** | **0.001 ± ~0** |
| | Test | AA-MMD[44] | <u>0.068 ± ~0</u> | <u>0.058 ± ~0</u> | <u>−21.782 ± 0.534</u> | <u>0.543 ± 0.004</u> |
| | | AA-MMD[44] + DenoiseNet | 0.031±~0 | 0.008±~0 | −28.723±0.501 | 0.010±0.001 |
| | | ID-UNet[28] | 0.009 ± ~0 | 0.006 ± ~0 | −34.476 ± 0.653 | 0.002 ± ~0 |
| | | ID-UNet[28] + DenoiseNet | 0.007 ± ~0 | 0.005 ± ~0 | −35.871 ± 0.632 | 0.002 ± ~0 |
| | | Triple-CBCT[32] | 0.006 ± ~0 | 0.001 ± ~0 | −41.778 ± 0.432 | 0.004 ± ~0 |
| | | SinoNet-FBP | 0.037 ± ~0 | 0.036 ± 0.001 | −28.724 ± 0.631 | 0.462 ± 0.003 |
| | | rFast-MMDNet | **0.002 ± ~0** | **~0 ± ~0** | **−55.708 ± 0.941** | **0.002 ± ~0** |
| *Fibroglandular* | Validation | AA-MMD[44] | <u>0.223 ± ~0</u> | <u>0.195 ± ~0</u> | <u>−10.753 ± 0.782</u> | <u>0.543 ± 0.002</u> |
| | | AA-MMD[44] + DenoiseNet | 0.017±~0 | 0.019±~0 | −27.452±0.701 | 0.012±0.006 |
| | | ID-UNet[28] | 0.014 ± ~0 | 0.017 ± ~0 | −31.001 ± 0.681 | 0.004 ± 0.004 |
| | | ID-UNet[28] +DenoiseNet | 0.012 ± ~0 | 0.014 ± ~0 | −32.178 ± 0.632 | 0.003 ± 0.001 |
| | | Triple-CBCT[32] | 0.011 ± ~0 | 0.004 ± ~0 | −35.343 ± 0.721 | 0.003 ± ~0 |
| | | SinoNet-FBP | 0.279 ± ~0 | 0.237 ± 0.016 | −9.631 ± 0.578 | 0.312 ± 0.003 |
| | | rFast-MMDNet | **0.008 ±~0** | **0.002 ±~0** | **−40.654 ± 0.591** | **0.002 ± ~0** |

| | Method | | | | |
|---|---|---|---|---|---|
| Test | AA-MMD[44] | $\underline{0.201 + \sim 0}$ | $\underline{0.186 + \sim 0}$ | $\underline{-12.457 + 0.889}$ | $\underline{0.482 + 0.004}$ |
| | AA-MMD[44] + DenoiseNet | $0.016 \pm \sim 0$ | $0.017 \pm \sim 0$ | $-28.525 \pm 0.619$ | $0.010 \pm 0.004$ |
| | ID-UNet[28] | $0.012 \pm \sim 0$ | $0.015 \pm \sim 0$ | $-32.375 \pm 0.794$ | $0.002 \pm 0.005$ |
| | ID-UNet[28] + DenoiseNet | $0.010 \pm \sim 0$ | $0.013 \pm \sim 0$ | $-33.785 \pm 0.513$ | $0.002 \pm 0.001$ |
| | Triple-CBCT[32] | $0.009 \pm \sim 0$ | $0.002 \pm \sim 0$ | $-36.784 \pm 0.741$ | $0.002 \pm \sim 0$ |
| | SinoNet-FBP | $0.264 \pm \sim 0$ | $0.229 \pm 0.016$ | $-11.619 \pm 0.682$ | $0.293 \pm 0.006$ |
| | rFast-MMDNet | $\mathbf{0.007 \pm \sim 0}$ | $\mathbf{0.001 \pm \sim 0}$ | $\mathbf{-43.349 \pm 0.692}$ | $\mathbf{0.001 \pm \sim 0}$ |

**Tab. 2**: The measured mean values and stds of difference between GTs and predictions of individual tissues. Values marked with a wavy underline are the weakest score in the validation set, those marked in the straight underline are the weakest scores in the test set. The strongest scores in validation and test sets are denoted in bold fonts.

### 3.3 Ablation Study

Ablation studies regarding the components of rFast-MMDNet and Triple-CBCT are conducted as seen in **Tab. 3**. Since the overall setup between Triple-CBCT and rFast-MMDNet is different, we slightly modified the network input/output channels per model adaption and keep the rest architecture unaltered. In specific, to adapt SD-Net and ID-Net into our rFast-MMDNet pipeline, we modified the output channel of SA-Net and the input channel SDNet from k (k=16) to 3. For adaption of SinoNet and DenoiseNet into Triple-CBCT framework, we changed the output channel of SinoNet and the input channel of DenoiseNet from 3 to 16. Since we propose to use loss combined with PSNR, SSIM, and $L_1$ deviation while Triple-CBCT used mean squared error (MSE) for model training, we also cross trained all the networks in **Tab. 3** with the two loss functions. We can observe from **Tab. 3** that models trained with combination loss can achieve better results than those trained with MSE loss. Addtionally, both SinoNet and DenoiseNet can perform marginally better than SD-Net as well as ID-Net in the current dataset.

| Method | Loss Function | RMSE | MAE | -PSNR | SSIM |
|---|---|---|---|---|---|
| SD-Net-FBP | Combination | $0.228 \pm 0.125$ | $0.153 \pm 0.109$ | $-12.422 \pm 1.452$ | $0.402 \pm 0.011$ |
| | MSE | $\underline{0.231 + 0.129}$ | $\underline{0.174 + 0.134}$ | $\underline{-10.205 + 1.878}$ | $\underline{0.487 + 0.014}$ |
| SinoNet-FBP | Combination | $\mathbf{0.173 \pm 0.095}$ | $\mathbf{0.120 \pm 0.095}$ | $\mathbf{-15.616 \pm 1.378}$ | $\mathbf{0.381 \pm 0.003}$ |
| | MSE | $0.187 \pm 0.103$ | $0.143 \pm 0.018$ | $-11.734 \pm 1.675$ | $0.404 \pm 0.007$ |
| SD-Net-FBP + DenoiseNet | Combination | $0.006 \pm 0.003$ | $0.006 \pm 0.006$ | $-41.078 \pm 0.498$ | $0.006 \pm \sim 0$ |
| | MSE | $0.006 \pm 0.005$ | $0.007 \pm 0.015$ | $-39.843 \pm 0.595$ | $0.007 \pm \sim 0$ |
| SinoNet-FBP + DenoiseNet (rFast-MMDNet) | Combination | $\mathbf{0.004 \pm \sim 0}$ | $\mathbf{0.005 \pm \sim 0}$ | $\mathbf{-45.027 \pm 0.542}$ | $\mathbf{0.002 \pm \sim 0}$ |
| | MSE | 0.006 ± 0.003 | 0.006 ± 0.007 | −42.108 ± 0.664 | 0.005 ± ~0 |
| SinoNet-FBP + ID-Net | Combination | 0.005 ± ~0 | 0.006 ± ~0 | −43.073 ± 0.587 | 0.004 ± ~0 |
| | MSE | 0.006 ± 0.004 | 0.007 ± 0.008 | −41.878 ± 0.547 | 0.006 ± ~0 |
| SD-Net-FBP + ID-Net (Triple-CBCT) | Combination | 0.006 ± 0.004 | 0.007 ± 0.008 | −40.032 ± 0.515 | 0.007 ± ~0 |
| | MSE | 0.007 ± 0.006 | 0.008 ± 0.015 | −38.571 ± 0.631 | 0.008 ± ~0 |

**Tab. 3**: Ablation Study of rFast-MMDNet with Triple-CBCT on test set. Combination loss is our proposed loss where MSE loss is the one used in Triple-CBCT framework. Comparison is made wthin stage-wise where SD-Net-FBP and SinoNet-FBP are in one comparison group and the rest is in the other. The best results are bolded and the weakest ones are underlined.

## 4. Discussion

Breast CT provides high-resolution 3D breast information non-invasively for various diagnostic tasks, which often require quantitive differentiation of tissues via multiple-material decomposition (MMD) [45]. Being an under-determined problem, the additional attenuation information presented in the second energy channel of DECT improves the accuracy and confidence of MMD for breast carcinoma diagnosis [45]. On the other hand, the additional information in DECT may still be inadequate to completely solve the MMD problem with more than two materials commonly involved in breast imaging and diagnosis. Thus far, existing >2 breast DECT MMD methods mostly decomposed on the reconstructed image using algebraic/physical-theory integrated hand-crafted models, which have been shown inaccurate, sensitive to domain shift, and slow to optimize for each incoming image [45–47]. A recent learning method on sinogram Triple-CBCT showed the feasibility of >2 material MMD but used a network structure that is difficult to train, generalize, and impose quality control. Therefore, we present a robust two-stage rFast-MMDNet trained directly on dual-energy CT projections for multiple material decompositions. The network comprises two 3D CNNs (SinoNet and DenoiseNet) trained independently for DE projection decomposition and image post-processing. SinoNet and DenoiseNet share the same loss objective combining negative PSNR, L1, and MS-SSIM with two tunable hyper-parameters to enforce model optimization. Forward and filtered back projections were also integrated into rFast-MMDNet for domain adaptation. Quantitative and qualitative evaluations were carried out with our pipeline on the 2022 AAPM DL-Spectral-CT dataset. Previously reported DECT MMD algorithms, including AA-MMD, AA-MMD + DenoiseNet, ID-UNet,

ID-UNet + DenoiseNet, and Triple-CBCT, were selected for comparison.  In both visual and numeric experimental results, rFast-MMDNet was competitive in all evaluation metrics and outperformed all comparison methods. Besides average error and similarity metrics, rFast-MMDNet resulted in the lowest and close to zero standard deviation, indicating superior model robustness.

We compared rFast-MMDNet with four representative methods, including hand-crafted image-domain direct inversion (AA-MMD), DL-based image-domain decomposition (ID-UNet), image domain MMD (AA-MMD/ID-UNet) + DenoiseNet, and DL based projection-domain MMD (Triple-CBCT) [32]. Unsurprisingly, without a learning component, AA-MMD performed worst, showing high noise level and poor boundary fidelity. Both AA-MMD and ID-UNet, which operated in the image domain, were hampered by the information loss in domain transfer.  Triple-CBCT and rFast-MMDNet hold a theoretical advantage due to their direct access to raw information in the sinograms. The theoretical advantage is confirmed by the superior performance of the two sinogram-based methods vs. image domain MMD + DenoiseNet methods on the breast data.

We attribute the superior performance of rFast-MMDNet in comparison to SinoNet + FBP to the synergy of SinoNet + DenoiseNet. Because the composition of a decomposed image ($I'$) in theory can be seen as information ($I$) + noise ($N$), both factors influence the predicted image quality. The difference between MMD in image and sinogram domains lies in $I$, where properly designed sinogram MMD theoretically can achieve better $I$ than that of image MMD. Accordingly, though in the case of rFast-MMDNet, FBP as an imperfect backprojection option that introduces streaking and cupping artifacts, those artifacts can be practically treated as noise and factored into $N$. Hence, it is mostly the artifacts introduced by FBP in $N$ that lead to to significantly worse performance of SinoNet + FBP than AA-MMD and ID-UNet.  Thus, we ran DenoiseNet right after the step of SinoNet + FBP to suppress $N$ from $I'$. Whereas, in image-domain MMD, $N$ is already suppressed in the reconstruction process with accompanying information loss from the sinogram space to the imaging space. In comparison, sinogram based MMD methods keep richer information and more noise. Subsequently as shown in Tab. 1 and Tab. 2, DenoiseNet improves SinoNet + FBP results more than those of AA-MMD and ID-UNet.

Of the two DL methods on sinograms, rFast-MMDNet is superior to Triple-CBCT due to the following attributes. 1) Clearer pipeline design to match network with task: Triple-CBCT splits the task of projection decomposition between SD-Net and ID-Net for transmission to VMI sinograms and VMI grouping, respectively, while multitasking on image refinement. The unclear task distribution among network modules limited model performance and interpretability. In contrast, rFast-MMDNet divides projection decomposition and image post-processing into separate SinoNet and DenoiseNet, each with its own objective. The clear separation reduces the training burden and encourages fast convergence. 2) More robust Network Design in two fold: First, the SD-Net in Triple-CBCT uses mixed spatial filter sizes of $1 \times 3$ and $3 \times 3$, and randomly plugs several skip-connection branches into the network flow, which causes inconsistent information to be passed throughout convolutions. Also, SD-Net consists of only seven layers, which limits the number of trainable parameters and model complexity. SinoNet, as the first stage of rFast-MMDNet, has a balanced U-shape with consistent skip-information flow between encoding and decoding. The balanced network combines 15 layers (7 encoding + 1 bottleneck + 7 decoding), with all layers having $3 \times 3$ convolution and $2 \times 2$ max-pooling filters, to achieve higher network learning and generalization capacity. Second, instead of directly stacking eight convolutional layers with inter-layer skip-connection for image domain processing in ID-Net of Triple-CBCT, DenoiseNet improves image denoising efficiency by stacking 16 ResNet blocks, where each ResNet Block has consistent intra-layer skip connection to ensure denser residual information feeding and effectively avoids gradient vanishing in backpropagation. 3) Reduced training load: Compared to the domain transfer net of Triple-CBCT that requires additional training, rFast-MMDNet relaxes the requirement of accurate domain adaption and leaves the final polish of the image quality to DenoiseNet. As a result, rFast-MMDNet requires the training of fewer networks while improving the prediction accuracy.

The top-ranked methods from the AAPM challenge achieved near-zero errors, further improvement of which is neither meaningful nor practical. While achieving comparable performance to these reported methods, we believe rFast-MMDNet holds a theoretical advantage in speed without incorporating the time-consuming iterative reconstruction as part of the network training and inference in most top performers [34].

Despite the encouraging results of automatic MMD frameworks demonstrated by rFast-MMDNet, they are several limitations. First, the supervision for SinoNet is forward projection sinograms, which are generated assuming ideal projection conditions. Specifically, a continuous object is discretized into a finite number of parameters per ray tracing modeling [48] for a fast and memory-efficient implementation of forward projection. Alternatively, more accurate and realistic sinograms can be generated using methods such as the pixel-basis/Kaiser-Besel basis functions or distance-driven forward projections [48], at a significantly higher computational cost. Second, FBP is used as the back projection algorithm in rFast-MMDNet for fast domain transformation, which could introduce unwanted artifacts in decomposed images. In the present solution, DenoiseNet is implemented to mitigate, not eliminate, the issue since limited data is available for model training. In future study, while more private/public data is available, we will consider combining transfer and federated learning to train a more robust and versatile denoise network through sufficient expose to diverse kinds of noise for better suppression of the artifacts introduced by FBP [49]. Thirdly, we used a synthesis dataset to evaluate rFast-MMDNet, where the GT is available and "noise-free". Yet often times GTs are missing or noisy in practice. Therefore, rFast-MMDNet needs to be further tested on real-world data. Fourthly, the proposed rFast-MMDNet requires raw material sinograms training supervision, which is hard to obtain without accurate geometry of the DECT scanner. Fifthly, we were unable to directly compare rFast-MMDNet with AAPM Grand Challenge Top Performers due to insufficient details in [34] to reproduce their work. The comparison is needed to quantify if the speed difference is relevant in the clinical setting. Last but not least, although rFast-MMDNet has an architecture conducive to generalization, the point needs to be further validated on additional datasets, particularly patient datasets acquired from different scanners and protocols. Currently, both our training and testing are limited by the available data. Data scarcity is a common problem for machine learning and deep learning research. Efforts have been made to increase the data availability via auto-/semi-auto annotation and mitigation of confidentiality concerns via techniques such as federated learning [50].

## 5. Conclusion

A robust and efficient two-stage DECT MMD pipeline to achieve an accurate decomposition of three materials directly using projection sinograms is presented in this study. The method was tested on digital breast phantom images with ground truth classification of the fibroglandular, adipose and calcification materials. Both qualitative and quantitative results show the advantage of using our network for fast, robust, and accurate MMD on DECT.